\documentclass[sigconf]{acmart}

\usepackage{color, colortbl}
\usepackage{enumitem}

\usepackage{amsmath, bm}

\setcopyright{acmcopyright}
\copyrightyear{2023}
\acmYear{2023} 
\setcopyright{acmlicensed}\acmConference[FAccT '23]{2023 ACM Conference on Fairness, Accountability, and Transparency}{June 12--15, 2023}{Chicago, IL, USA}
\acmBooktitle{2023 ACM Conference on Fairness, Accountability, and Transparency (FAccT '23), June 12--15, 2023, Chicago, IL, USA}
\acmPrice{15.00}
\acmDOI{10.1145/3593013.3593985}
\acmISBN{979-8-4007-0192-4/23/06}

\begin{document}


\title{WEIRD FAccTs: How~Western,~Educated, Industrialized, Rich,~and~Democratic~is~FAccT?}


\author{Ali Akbar Septiandri, Marios Constantinides, Mohammad Tahaei, Daniele Quercia}
\email{{ali.septiandri, marios.constantinides, mohammad.tahaei, daniele.quercia}@nokia-bell-labs.com}
\affiliation{%
  \institution{Nokia Bell Labs}
  \city{Cambridge}
  \country{UK}
}


\begin{abstract}
Studies conducted on \textbf{W}estern, \textbf{E}ducated, \textbf{I}ndustrialized, \textbf{R}ich, and \textbf{D}emocratic (WEIRD) samples are considered atypical of the world's population and may not accurately represent human behavior. In this study, we aim to quantify the extent to which the ACM FAccT conference, the leading venue in exploring Artificial Intelligence (AI) systems' fairness, accountability, and transparency, relies on WEIRD samples. We collected and analyzed 128 papers published between 2018 and 2022, accounting for 30.8\% of the overall proceedings published at FAccT in those years (excluding abstracts, tutorials, and papers without human-subject studies or clear country attribution for the participants). We found that 84\% of the analyzed papers were exclusively based on participants from Western countries, particularly exclusively from the U.S. (63\%). Only researchers who undertook the effort to collect data about local participants through interviews or surveys added diversity to an otherwise U.S.-centric view of science. Therefore, we suggest that researchers collect data from under-represented populations to obtain an inclusive worldview. To achieve this goal, scientific communities should champion data collection from such populations and enforce transparent reporting of data biases.
\end{abstract}



\begin{teaserfigure}
  \includegraphics[width=\textwidth]{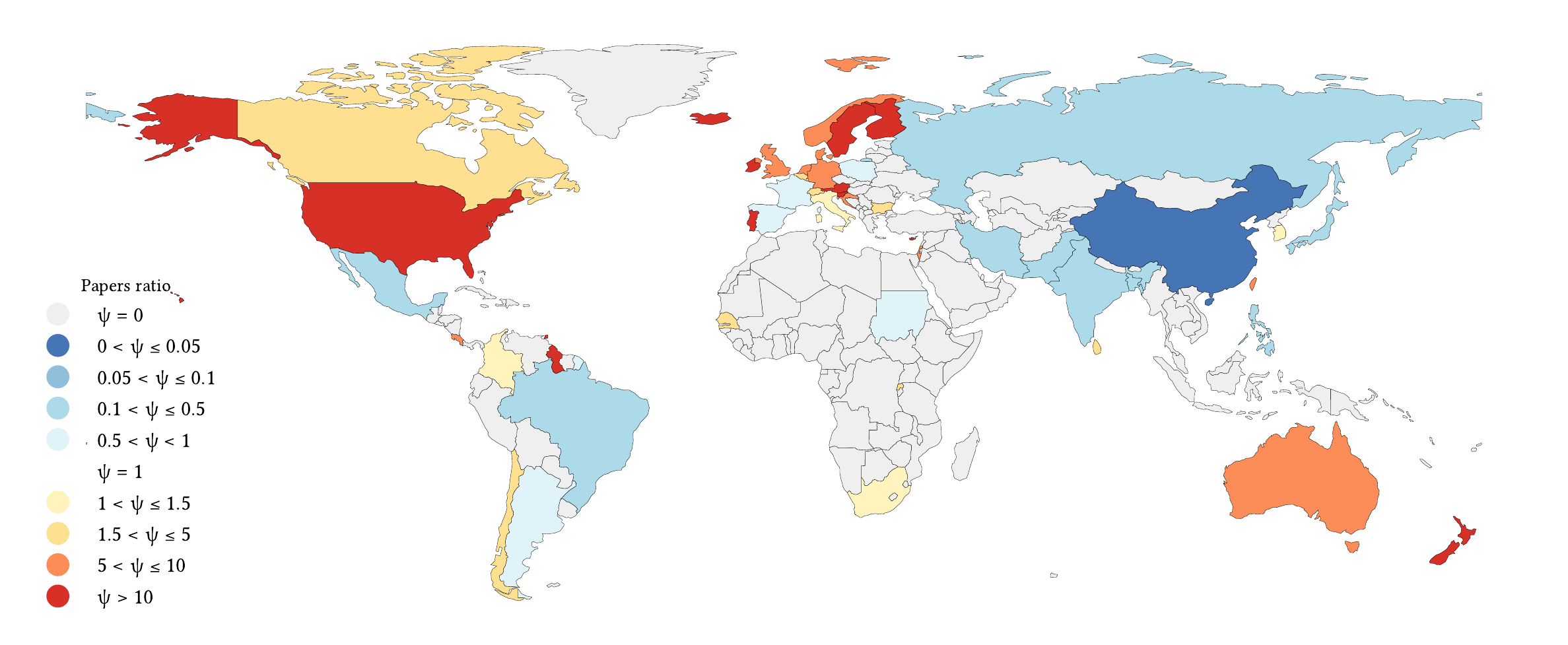}
  \caption{Worldwide distribution of FAccT papers ratio ($\psi_{sc}$) between 2018--2022, showing over-represented ($\psi_{sc} > 1$) and under-represented countries ($\psi_{sc} < 1$). The papers ratio $\psi_{sc}$ is the number of papers from a given country $c$ (where the study participants are from) over the number of all papers in FAccT, divided by $c$'s population over the world's population. Countries in light gray ($\psi_{sc} = 0$) did not have participants in FAccT between 2018-2022, while those in darker shades of blue and red indicate under- and over-represented countries, respectively.}
  \label{fig:map-facct}
\end{teaserfigure}



\maketitle

\section{Introduction}
\label{sec:introduction}
The ACM Conference on Fairness, Accountability, and Transparency (FAccT)\footnote{Established in 2018, and initially called FAT.} is an annual conference that aims to bring together a diverse community of scholars interested in exploring the fairness, accountability, and transparency of Artificial Intelligence (AI) and Machine Learning (ML) systems. One of the FAccT's challenges concerns biases, particularly as AI is increasingly used in decision-making contexts~\cite{elliot2021state}. One way biases can manifest is through the overuse of specific datasets when evaluating ML models, leading to unfair or inaccurate results. For example, ImageNet, a popular dataset for computer vision tasks, was found to be biased~\cite{yang2020imagenet, steed2021imagenet}. 

In recent years, there has also been increasing attention on the concept of \textbf{W}estern, \textbf{E}ducated, \textbf{I}ndustrialized, \textbf{R}ich, and \textbf{D}emocratic (WEIRD) research, which refers to studies conducted on participants drawn from WEIRD populations~\cite{henrich2010weirdest}. These samples are often considered atypical of the global population and may not represent human behavior more globally. For example, an analysis of the Conference on Human Factors in Computing Systems (CHI), the prominent conference for empirical studies focused on human-centered computing, showed that 73\% of CHI papers come from Western samples, representing less than 12\% of the world's population~\cite{linxen2021weird}. Such a focus on WEIRD samples results in the design of technologies used by the rest of the population. However, it does not consider those populations' cultural norms, characteristics, and expectations~\cite{niess2021attitudes, sambasivan2018privacy}.

Given the potential biases introduced by WEIRD samples, it is therefore vital to consider the extent to which a conference's published papers rely on such samples. In this study, we are interested in quantifying the \emph{WEIRD-ness} of the FAccT conference by examining the datasets used in its proceedings, following \citet{linxen2021weird}'s method for analyzing CHI's WEIRD-ness. FAccT is a noteworthy conference to study WEIRD-ness as one of its main objectives is to bring fairness, accountability, and transparency to AI. In so doing, we made three main contributions:

\begin{itemize}
    \item We collected the FAccT proceedings between 2018 and 2022, resulting in an initial set of 416 papers. Out of these papers, we analyzed a total of \textbf{128 papers} that met the inclusion criteria for our analysis (\S\ref{sec:methodology}).\footnote{We made our dataset publicly available for replication and reproducibility.}
    \item We found that \textbf{84\%} of the analyzed papers were exclusively based on participants from Western countries (\S\ref{sec:results}).
    \item Compared to CHI, FAccT has a more significant proportion of Western study participants (\S\ref{sec:results}). By analyzing the intersection of the study participants from the two conferences, we found that the difference in ``EIRD'' (i.e., Educated, Industrialized, Rich, and Democratic) between the two conferences is still significant. To explain these differences, we looked at the sample size, types of datasets used (off-the-shelf vs. author-collected), and authors' affiliation country. However, we did not find evidence to attribute these differences to any of these factors.
\end{itemize}

In light of these results, we discuss the theoretical and practical implications of our findings (\S\ref{sec:discussion}), which speak to the need to take steps such as diversifying authors, encouraging collaborations through workshops, and using cross-cultural online research platforms to make FAccT, and more broadly computing conferences, less WEIRD.

\section{Related Work}
\label{sec:related}


In 2010, \citeauthor{henrich2010weirdest} did a comprehensive study about the generalizability of research findings. They state that most research findings are based on a small world population, often called the WEIRD population. In psychology, 96\% of research samples come from the WEIRD population, which only accounts for 12\% of the world population. One reason for such bias toward WEIRD samples is that many authors (73\%) come from American universities. The situation exacerbates when authors and papers often assume and claim that their findings are universally valid and generalizable, asserting applicability to ``humans.'' Nonetheless, the study population sometimes barely goes beyond undergraduate students from the authors' home institute~\cite{henrich2010weirdest}. Additionally, studies have demonstrated an unequal distribution of participation in the AI ethics debate~\cite{jobin2019globalai}. There is also a lack of awareness regarding the potential consequences of AI adoption across nations, such as the adoption of face recognition technology in public administration in India despite known biases associated with women and minorities~\cite{sengupta2023globalsouth}.

Since then, in the past decade, there has been an emerging interest in the Computer Science community about the potential biases introduced by WEIRD samples. A few examples of such research that informs the biases introduced by WEIRD samples include the consideration of: cultural differences when designing products that are used in many countries across the globe~\cite{niess2021attitudes, busse2020cash, wilkinson2022many, ma2022enthusiasts}, minorities in the design of computing education materials~\cite{oleson2022decade}, and countries in the Global South for the design of digital accessibility~\cite{nourian2022digital}. To shed light on these biases, we provide details for two studies: (1) In the case of personal trackers, the goals of different user types can vary. Arab users generally see a fitness tracker more as a monitoring device. In contrast, WEIRD populations have been found to see the same device as an assistance to achieve their goal. In these two scenarios, a monitoring device would give users advice and recommendations, while a tool with more authority acts as a coach and gives rules and guidelines to the user~\cite{niess2021attitudes}; (2) Smartphone privacy is influenced by the users' cultural norms. The current design may assume that device sharing does not happen, and privacy is preserved as long as a phone lock feature is provided. However, in some countries (e.g., India, Pakistan, and Bangladesh), users, particularly female users, may not be able to exercise their autonomy. It may be considered socially unacceptable not to let other people use their phones. In such scenarios, giving deletion options and private mode browsing could be options to help users maintain their privacy and adhere to their social norms (by letting other people use their phone)~\cite{sambasivan2018privacy}.

    In particular, the two computing conferences that heavily focus on sampling issues are \textit{CHI}, which focuses on human-centered design, and \textit{FAccT}, which aims to democratize AI and advance the development of responsible AI. There has been a recent meta-study on CHI findings from 2016--2020 about the WEIRD-ness of the conference (\S\ref{sec:rw-original-study}), but not on FAccT. \citeauthor{laufer2022four} conducted a meta (reflexive) study on four years of FAccT proceedings to extract research topics (e.g., group-level fairness and disinformation) and understand community's values (e.g., transparency over the peer review process and industry influence over the published research)~\cite{laufer2022four}. Our work contributes to this line of research by exploring FAccT's WEIRD-ness, and comparing it with another prominent research conference that focuses on empirical research and human-centered computing (CHI).

\textbf{The Original Study: How WEIRD Is CHI?}
\label{sec:rw-original-study}
\citet{linxen2021weird} collected 3,269 CHI papers published in 2016--2020. Of these papers, 2,768 were on human subjects, and 1,076 papers (38.9\%) had information about the participants' countries, which is the basis of the study. CHI findings span 93 countries, with 7.1\% of the samples being from Western countries. The USA, Ireland, and Switzerland were the top three represented countries regarding the number of participants. The correlation between samples and other EIRD factors was also positive; the higher these factors, the stronger those populations were presented in the dataset. Similar to what \citeauthor{henrich2010weirdest} suggested (diversifying authors could be a way to reduce WEIRD research), findings from CHI show that 81.2\% of papers studied participants from the authors' institute's country. The authors propose ideas such as diversifying authorship, increasing the use of online crowd-sourcing platforms, and appreciating replication studies to make CHI less WEIRD.
\section{Methodology}
\label{sec:methodology}
Based on our related work (\S\ref{sec:related}), we set out to explore the WEIRD-ness of FAccT studies. In so doing, we formulated two Research Questions (RQs): \textbf{RQ\textsubscript{1}}: How WEIRD is the FAccT conference? \textbf{RQ\textsubscript{2}:} How FAccT and CHI conferences compare in terms of WEIRD-ness? To answer these RQs, we replicated the \citet{linxen2021weird}'s methodology. \citeauthor{linxen2021weird}'s research was the first to operationalize WEIRD-ness in a conference proceeding, and by comparing FAccT with CHI, we demonstrate similarities and differences between the two communities and surface potential areas of oversight. We collected and analyzed 416 papers published at FAccT conferences between 2018--2022 from the ACM Digital Library.



\subsection{Positionality Statement}
Understanding researcher positionality is essential to demystifying our lens on data collection and analysis~\cite{frluckaj2022gender, havens2020situated}. We situate this paper in the United Kingdom in the 21\textsuperscript{st} century, writing as authors who primarily work as academic and industry researchers. We identify as males from Indonesia, Cyprus, Iran, and Italy. Our shared backgrounds include HCI, privacy, security, software engineering, AI, social computing, and urbanism.

\subsection{Dataset}
The first author manually reviewed a random set of 103 (24.8\%) papers relevant to the RQs. From this review, a coding guideline and exclusion criteria were collaboratively developed with three other authors to extract relevant data from the papers. The coding scheme and the exclusion criteria were then reviewed and discussed by all authors to ensure their completeness and accuracy. To augment the information obtained from the papers, we emailed 42 dataset owners requesting any missing information. Out of these, 19 (45.2\%) replied, but we still could not get the country attribution for 9 (47.4\%) datasets. The collected data were then analyzed using the agreed-upon coding scheme. Overall, this multi-step process aimed at systematically and rigorously preparing the dataset and reducing potential biases. To factor in the true magnitude of participants in the statistical analyses, we multiplied the number of datasets that were used more than once by the number of papers using those datasets. Consider, for example, the German credit dataset with 1,000 participants \cite{hofmann1994german}. This dataset was used in 13 papers, yielding a total of 13,000 participants. This approach emphasized the problem in frequently used datasets, and aimed at motivating the community to consider under-represented countries in future sampling. We opted-in for multiplication because otherwise overused datasets would not show their magnitude in the analysis. For example, the German credit dataset would have been counted only once despite being used 13 times.

\subsection{Inclusion \& Exclusion Criteria}
\label{sec:inclusion-criteria}
We manually analyzed 416 papers published at FAccT conferences between 2018 and 2022 and found that 93 (22.36\%) of the papers used off-the-shelf datasets instead of manually collecting them through interviews, surveys, or workshops. To differentiate the two types of datasets, we will refer to these manually collected datasets from this point onward as ``author-collected datasets.'' For our analysis, we define our exclusion criteria as follows:
\begin{itemize}
    \item \textit{Abstracts and tutorials}: No study or insufficient information about the described studies;
    \item \textit{Non-human datasets}, e.g., MNIST (handwritten digit images), CIFAR-10 (tiny images of 10 objects), Stanford Dogs (images of 120 breeds of dogs), ImageNet (14 million images of more than 20,000 categories), Singapore Bus Stops~\cite{tedjopurnomo2022equitable}, Boston Housing~\cite{harrison1978hedonic}: because mixing human and non-human data would make it difficult to compare the results, as the units of measurement would be different;
    \item \textit{Country attribution is unclear}, e.g., OpenImages MIAP, Last.FM, Amazon reviews: because it was not clear which country the human subjects were from or because the authors were studying issues related to a specific country but not analyzing data from human subjects.
\end{itemize}

\subsection{Defining WEIRD Scores}
The goal of this work is to quantify the WEIRD-ness of FAccT. To achieve that, we defined a set of WEIRD scores relying on two strands of research: \emph{a)} work that focused on certain populations of the CHI conference~\cite{linxen2021weird}; and, \emph{b)} work that focused on specific products for particular  populations without much consideration for a majority of the world's population~\cite{niess2021attitudes, nourian2022digital}. Next, we explain each of the WEIRD scores (Table~\ref{tab:formulae}): \\

\noindent \textbf{Western.} To determine whether a country is classified as Western, we used the Huntington classification~\cite{huntington2000clash}. This is based on Huntington's thesis, \textit{the Clash of Civilizations}, where he posits that cultural differences are the main cause of conflicts among humanity. Therefore, the classification is based on cultural and historical factors, including the country's language and religion. According to Huntington, Western civilization is rooted in the tradition of ancient Greece and Rome, and is characterized by certain commonalities such as the predominance of Christianity, the use of the Latin alphabet, and the prevalence of democratic political systems. Of particular interest in his thesis is the concept of ``torn countries''; countries with no clear-cut classification. For example, Turkey has adopted Western customs, including how its people dress, started using the Latin alphabet, became a member of NATO, and has been trying to become a member of the European Union. However, under Huntington's classification, Turkey is still classified as non-Western due to its history, culture, and traditions derived from Islamic civilization. All European Union members are classified as Western countries~\cite{eu2022country}. It is worth mentioning that certain nations, such as Japan, South Korea, Chile, and Argentina, meet the criteria of being Educated, Industrialized, Rich, and Democratic, but do not fall under the category of Western nations.

\noindent \textbf{Educated.} We used the mean years of schooling per person as reported in the UNDP Human Development Report~\cite{undp2022hdr} to measure the educational level of a country. This measure considers the mean years of schooling for adults aged 25 years. An alternative proxy would be the PISA index from OECD. However, the UNDP index was favored to allow for reproducible results (similar to CHI's findings~\cite{linxen2021weird}).

\noindent \textbf{Industrialized.} To determine the industrialization of a country, we used the gross domestic product (GDP) per capita adjusted by purchasing power parity (PPP)~\cite{worldbank2022gdp}. GDP measures the total value of goods and services produced within a country's borders in a given period, usually a year. It is an indicator of a country's economic performance and is used to compare the economic output of different countries. The data is reported in current international dollars. The international dollar is an artificial currency used to adjust for differences in purchasing power when comparing economic performance among countries. An alternative to this proxy is to use the Competitive Industrial Performance (CIP) Index by United Nations Industrial Development Organization (UNIDO) \cite{cipi2020unido}. However, to produce comparable results to CHI's WEIRD-ness, the choice of GDP per capita adjusted by PPP was favored.

\noindent \textbf{Rich.} We used the gross national income (GNI) per capita adjusted by purchasing power parity (PPP)~\cite{worldbank2022gni} to measure the wealth of a country as suggested by \citet{arnett2016neglected} and \citet{linxen2021weird}. GNI per capita approximates a population's standard of living as it is calculated by adding up all the income earned by a country's residents and businesses. The data is reported in current international dollars, allowing for comparison between countries.

\noindent \textbf{Democratic.} To determine the democracy of a country, we used the ``political rights'' scores provided by the Freedom House~\cite{freedom2022countries}. Freedom House is a U.S. non-profit organization whose mission is to research democracy, freedom, and human rights. The ``political rights'' measure considers the level of political freedom and rights enjoyed by the citizens of a country. Alternatively, one could use the Democracy Index \cite{democracy2022economist} as a proxy for political rights,
developed by the research division of the Economist Group, the Economist Intelligence Unit (EIU). To allow for reproducible results, we opted-in for political rights provided by the Freedom House.

\subsection{Analysis: Terminology and Computing WEIRD}

\begin{table*}
    \centering
    \caption{WEIRD variables and the formula to compute the values. $\mathbb{E}[.]$ means the expected value of a random variable. In Kendall rank correlation ($\tau$), $P$ is the number of concordant pairs, $Q$ is the number of discordant pairs, $T$ is the number of ties in the first variable, and $U$ is the number of ties in the second variable. Concordant pairs are pairs of observations in which the two variables are ranked in the same order, while discordant pairs are pairs of observations in which the two variables are ranked in opposite orders~\cite{agresti2010analysis}. $\vec{X}$ means a vector of value for variable $X$ from all sampled countries. The formulae in this table are derived from and formalized based on the methods outlined in the paper by \citet{linxen2021weird}.}
    \label{tab:formulae}
    \begin{tabular}{lllp{7cm}}
        \toprule
        Symbol & Variable & Formula & Description \\
        \midrule
        $c$ & Country & - & Country where the samples are from \\
        $p$ & Participants & - & Number of participants \\
        $s$ & Papers & - & Number of papers \\
        $\pi$ & Population & - & Population size of a country based on World Population Prospects 2022 \cite{population2022un} \\
        \midrule
        $W_c$ & Western & $1 \text{ if } c \in \text{ Western else } 0$ & Whether country $c$ is Western based on Huntington classification~\cite{huntington2000clash} \\
        $E_c$ & Educated       & $\mathbb{E}_c[\text{years of schooling}]$ & Mean years of schooling for country $c$ based on UNDP Human Development Report (2022)~\cite{undp2022hdr} \\
        $I_c$ & Industrialized & $\text{GDP per capita}_c$ & Level of industrialization for country $c$ based on World Bank GDP per capita, PPP (current Int\$, 2020)~\cite{worldbank2022gdp} \\
        $R_c$ & Rich           & $\text{GNI per capita}_c$ & Wealth of country $c$ based on World Bank GNI per capita, PPP (current Int\$, 2020)~\cite{worldbank2022gni} \\
        $D_c$ & Democratic     & $\text{political rights}_c$ & Level of democracy for country $c$ based on Freedom House Political Rights (2022)~\cite{freedom2022countries} \\
        \midrule
        $\psi_{pc}$ & Participants ratio per country & $\frac{p_c / \sum_c p_c}{\pi_c / \sum_c \pi_c}$ & Ratio of the proportion of participants for country $c$ to the proportion of population size for country $c$ \\
        $\psi_{sc}$ & Papers ratio per country & $\frac{s_c / \sum_c s_c}{\pi_c / \sum_c \pi_c}$ & Ratio of the proportion of papers for country $c$ to the proportion of population size for country $c$ \\
        $\tau(., .)$ & Kendall rank correlation & $\frac{P - Q}{\sqrt{(P + Q + T) \cdot (P + Q + U)}}$ & The similarity of two rankings, e.g. $\vec{\psi}_s$ and $\vec{E}$; $\vec{\psi}_p$ and $\vec{D}$ \\
        \midrule
        $W$-score & Western score & $\frac{1}{N} \sum_c W_c$ & Expected value of how Western a conference is from all sampled countries \\
        $E$-score & Educated score & $\tau(\vec{\psi}_s, \vec{E})$ & How correlated papers ratio and mean years of schooling from all sampled countries \\
        $I$-score & Industrialized score & $\tau(\vec{\psi}_s, \vec{I})$ & How correlated papers ratio and level of industrialization from all sampled countries \\
        $R$-score & Rich score & $\tau(\vec{\psi}_s, \vec{R})$ & How correlated papers ratio and level of wealth from all sampled countries \\
        $D$-score & Democratic score & $\tau(\vec{\psi}_s, \vec{D})$ & How correlated papers ratio and level of democracy from all sampled countries \\
        \bottomrule
    \end{tabular}
\end{table*}

Throughout the paper, we refer to Table~\ref{tab:formulae} for the quantification of the WEIRD scores. The formulae are based on and formalized from methods in \cite{linxen2021weird}. Kendall rank correlation is often used when the data is not normally distributed or when the variables being compared are ordinal rather than interval or ratio variables~\cite{agresti2010analysis}. The coefficient ranges from -1 to 1, where a value of 1 indicates a perfect positive correlation, a value of -1 indicates a perfect negative correlation, and a value of 0 indicates no correlation. If the coefficient is close to 1, then the researchers at FAccT conferences fully conducted studies with WEIRD subjects. To get the Kendall rank correlation coefficient closer to zero, researchers should aim to make their study participants proportionate to the country's population~\cite{agresti2010analysis}.
\section{Results}
\label{sec:results}

We reviewed 416 papers and discovered that 226 (54.3\%) included either synthetic, off-the-shelf, or author-collected datasets. We further narrowed down the selection by excluding papers that were based on non-human participants, had an unspecified number of participants, or did not specify the country of origin, resulting in \textbf{128} papers (30.8\%) for analysis (\S\ref{sec:inclusion-criteria} for details of our (ex)in-clusion criteria).

\begin{figure}
    \centering
    \includegraphics[width=\columnwidth]{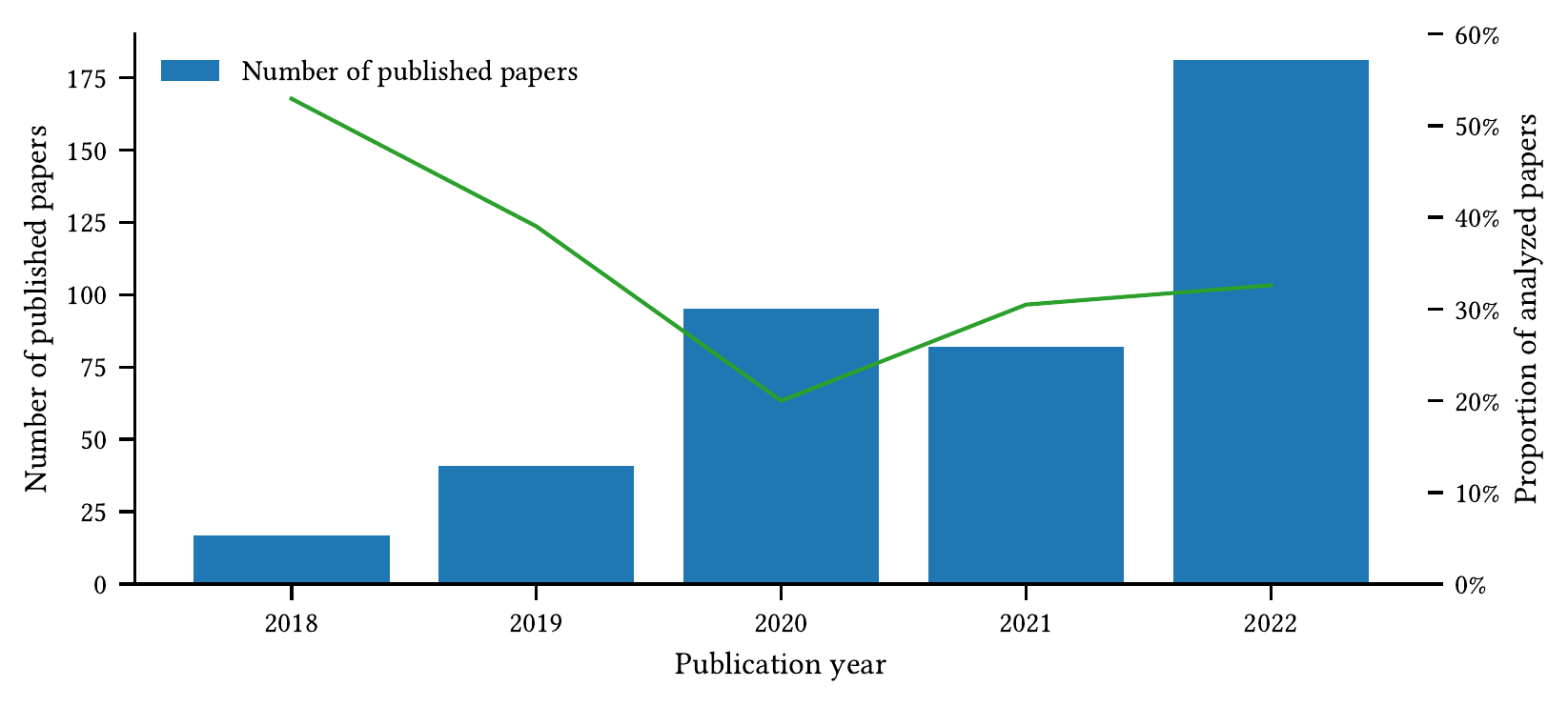}
    \caption{Number of published papers and proportions (green line) of those being analyzed. The proportion of the analyzed papers varied between years due to the number of published papers and the type of those papers. For example, in 2020, 36\% of the published papers were abstracts or tutorials. Tutorial papers were not archived in publication years other than 2020.}
    \label{fig:nb_papers}
\end{figure}

From 2018 to 2022, we observed an increase in the number of published papers in FAccT (Figure~\ref{fig:nb_papers}). In recent years, there have also been more papers with human participants (32.6\% in 2022). We saw this increase after the drop in 2020. The proportion of analyzed papers in 2020 was the lowest because 36\% of the published papers that year were either abstracts or tutorials; it was the only year when the tutorials were archived.

Table~\ref{table:top_countries} shows the top 10 countries by the participants ratio ($\psi_{pc}$), papers ratio ($\psi_{sc}$), and number of papers ($n_{sc}$). We found that the U.S. participants are over-represented in the FAccT community (93.6\% of the participants). This is because some datasets (e.g., Adult~\cite{uci2017ml}, COMPAS recidivism~\cite{larson2023compas}, and ACS~\cite{american2021, ding2021retiring}) are used multiple times and are based on national surveys. For similar reasons, Germany is the second highest country in terms of the number of papers because of the German credit dataset~\cite{uci2017ml} that is used in 13 (10.2\%) papers. 

\begin{table*}
    \caption{Top 10 countries of FAccT participants between 2018--2022. $\psi_{pc}$ is the participants ratio, $\psi_{sc}$ is the papers ratio, $n_{pc}$ is the number of participants, $n_{sc}$ is the number of papers, $\%_{pc}$ is the proportion of participants, and $\%_{sc}$ is the proportion of papers. The USA has the most number of papers and participants. One explanation for this is the availability of off-the-shelf datasets such as Adult, COMPAS recidivism, and ACS. These datasets, particularly Adult and ACS, are based on national surveys and, as a result, have many participants. Since the papers ratio $\psi_{sc}$ (middle-column) is calculated by the proportion of papers over the proportion of population size in country $c$, countries with smaller population size (e.g., Iceland, Guyana, Cyprus) are over-represented.}
    \label{table:top_countries}
    \resizebox{\textwidth}{!}{%
    \begin{tabular}{lrrr|lrrr|lrrr}
        \toprule
        \multicolumn{4}{c}{Top countries by $\psi_{pc}$} & \multicolumn{4}{c}{Top countries by $\psi_{sc}$} & \multicolumn{4}{c}{Top countries by $n_{sc}$} \\
        \midrule
        Country & $n_{pc}$ & $\%_{pc}$ & $\psi_{pc}$ & Country & $n_{sc}$ & $\%_{sc}$ & $\psi_{sc}$ & Country & $n_{sc}$ & $\%_{sc}$ & $\psi_{sc}$ \\
        \midrule
        USA & 16,621,168 & 93.62\% & 21.85 & Iceland & 1 & 0.44\% & 93.79 & USA & 106 & 46.49\% & 10.85 \\
        Colombia & 472,000 & 2.66\% & 4.09 & Guyana & 1 & 0.44\% & 43.14 & Germany & 20 & 8.77\% & 8.25 \\
        Portugal & 87,963 & 0.50\% & 3.77 & Ireland & 4 & 1.75\% & 27.81 & United Kingdom & 12 & 5.26\% & 6.15 \\
        Taiwan & 120,000 & 0.68\% & 2.22 & Cyprus & 1 & 0.44\% & 27.79 & India & 8 & 3.51\% & 0.20 \\
        Costa Rica & 22,000 & 0.12\% & 1.90 & Portugal & 7 & 3.07\% & 23.38 & Portugal & 7 & 3.07\% & 23.38 \\
        India & 399,528 & 2.25\% & 0.13 & Trinidad and Tobago & 1 & 0.44\% & 22.65 & Taiwan & 4 & 1.75\% & 5.77 \\
        Germany & 15,750 & 0.09\% & 0.08 & New Zealand & 3 & 1.32\% & 20.38 & Canada & 4 & 1.75\% & 3.63 \\
        Iceland & 63 & 0.00\% & 0.08 & Slovenia & 1 & 0.44\% & 16.24 & Sweden & 4 & 1.75\% & 13.27 \\
        Slovenia & 336 & 0.00\% & 0.07 & Sweden & 4 & 1.75\% & 13.27 & Ireland & 4 & 1.75\% & 27.81 \\
        Sweden & 1,471 & 0.01\% & 0.06 & Finland & 2 & 0.88\% & 12.44 & Australia & 4 & 1.75\% & 5.36 \\
        \bottomrule
    \end{tabular}
    }
\end{table*}

\subsection{RQ1: How WEIRD is the FAccT Conference?}
\label{sec:result-rq1}
Table~\ref{table:western} shows that the FAccT community primarily selects participants from Western countries (84\% exclusively from Western countries, 63\% exclusively from the U.S.). We observed that the proportion of FAccT papers with exclusively Western participants is higher than CHI. The classification of participant countries in a paper is determined by whether the participants are from Western or non-Western countries as defined by Huntington's framework~\cite{huntington2000clash}. Papers are considered: ``exclusively Western,'' if all participants are from Western countries; ``exclusively non-Western,'' if all participants are from non-Western countries; and, ``mixed,'' if participants are from both Western and non-Western countries. The data in the CHI column is not included in the paper by \citet{linxen2021weird} but for comparison, we reproduced the numbers using their data. In the paper by \citet{linxen2021weird}, the authors reported the number of ``participant samples'' since a single paper can report participants from multiple countries. As the reported number can be higher than the actual number of papers, our method matches the total number of papers.

\begin{table}
    \caption{Distribution of papers by participant countries. Most FAccT \& CHI papers use participants exclusively from Western countries.}
    \label{table:western}
    \begin{tabular}{lrr|rr}
        \toprule
        {} & \multicolumn{2}{c}{\textbf{FAccT}} & \multicolumn{2}{c}{\textbf{CHI} (\citeauthor{linxen2021weird})} \\
        Variable      &  n   & \%      &  n    &      \% \\
        \midrule
        Exclusively Western     &  108 & 84.38\% &  817 & 75.93\% \\
        Exclusively non-Western &    9 &  7.03\% &  197 & 18.31\% \\
        Mixed                   &   11 &  8.59\% &   62 &  5.76\% \\
        \midrule
        Total                   &  128 & 100\%   & 1076 & 100\%   \\
        \bottomrule
    \end{tabular}
\end{table}

Table~\ref{table:weird_index} shows the Kendall rank correlations ($\tau$) between EIRD values and the ratio of the paper ($\psi_{sc}$). We provided the 95\% Confidence Interval (CI) via bootstrapping.\footnote{Bootstrapping is a statistical method that involves resampling a dataset with replacement to quantify the uncertainty associated with a given estimator. This method is beneficial when the population distribution is unknown or when the sample size is small~\cite{james2013introduction}.} We also show the results from \citet{linxen2021weird} in the same table as a comparison point. As a sanity check, we reproduced their findings using our code and method\footnote{Our code repository and dataset are publicly available at \url{https://github.com/aliakbars/weird-facct}.}, and their dataset. According to Table~\ref{table:top_countries}, the percentage of participants from the U.S. in FAccT (93.6\%) is higher than in CHI (54.8\%). Nevertheless, Table~\ref{table:weird_index} suggests that the level of ``EIRD-ness'' is lower for FAccT. This may come from the EIRD scores---calculated using Kendall tau \emph{rank correlation}---which are not affected by outliers. For example, if half of the participants from the U.S. were removed, the proportion of the U.S. participants compared to all participants and the participant ratio would change. However, the U.S. would still rank first, and the ranks of other countries would remain unchanged. As a result, the EIRD scores would not change.

\begin{table}
    \caption{Kendall rank correlations ($\tau$) of the ratio of the paper $\psi_{sc}$ with measures of Educated, Industrialized, Rich, Democratic. The confidence intervals are generated from 10,000 bootstrap samples. Significance level: *$p < .05$, **$p < .01$, ***$p < .001$.}
    \label{table:weird_index}
    \begin{tabular}{lll|ll}
        \toprule
        {} & \multicolumn{2}{c}{\textbf{FAccT}} & \multicolumn{2}{c}{\textbf{CHI} (\citeauthor{linxen2021weird})} \\
        Variable       &  $\tau$ &   95\% CI $\tau$ &  $\tau$ &   95\% CI $\tau$ \\
        \midrule
        Educated       &  0.31**  &   [0.12, 0.50] & 0.46*** &   [0.34, 0.59] \\
        Industrialized &  0.35*** &   [0.21, 0.50] & 0.50*** &   [0.40, 0.62] \\
        Rich           &  0.35*** &   [0.20, 0.50] & 0.50*** &   [0.39, 0.62] \\
        Democratic     &  0.36*** &   [0.19, 0.53] & 0.50*** &   [0.38, 0.62] \\
        \bottomrule
    \end{tabular}
\end{table}

\subsection{RQ2: How FAccT and CHI Conferences Compare in Terms of WEIRD-ness?}
\label{sec:rq2}
In what follows, we sought to find possible explanations for these differences. To explain them, we examined three potential factors for the variations between FAccT and CHI: \emph{1)} we aimed to understand whether the sample size of CHI could account for the differences; \emph{2)} we evaluated the connection between the datasets used and the EIRD scores, only testing this hypothesis on FAccT as the datasets were only coded for this conference; \emph{3)} we evaluated the relationship between cross-country collaboration and the EIRD scores. 

\begin{table}
    \caption{Kendall rank correlations ($\tau$) of the ratio of the paper $\psi_{sc}$ with measures of Educated, Industrialized, Rich, Democratic by using the same set of countries for both FAccT and CHI. The confidence intervals are generated from 10,000 bootstrap samples. Significance level: *$p < .05$, **$p < .01$, ***$p < .001$.}
    \label{table:weird_intersection}
    \begin{tabular}{lll|ll}
        \toprule
        {} & \multicolumn{2}{c}{\textbf{FAccT}} & \multicolumn{2}{c}{\textbf{CHI} (\citeauthor{linxen2021weird})} \\
        Variable       &  $\tau$ &   95\% CI $\tau$ &  $\tau$ &   95\% CI $\tau$ \\
        \midrule
        Educated       &  0.40*** &   [0.18, 0.62] & 0.54*** &   [0.37, 0.72] \\
        Industrialized &  0.44*** &   [0.27, 0.60] & 0.65*** &   [0.55, 0.75] \\
        Rich           &  0.44*** &   [0.27, 0.60] & 0.65*** &   [0.55, 0.75] \\
        Democratic     &  0.45*** &   [0.27, 0.63] & 0.58*** &   [0.40, 0.76] \\
        \bottomrule
    \end{tabular}
\end{table}

\begin{figure*}
    \centering
    \includegraphics[width=.92\linewidth]{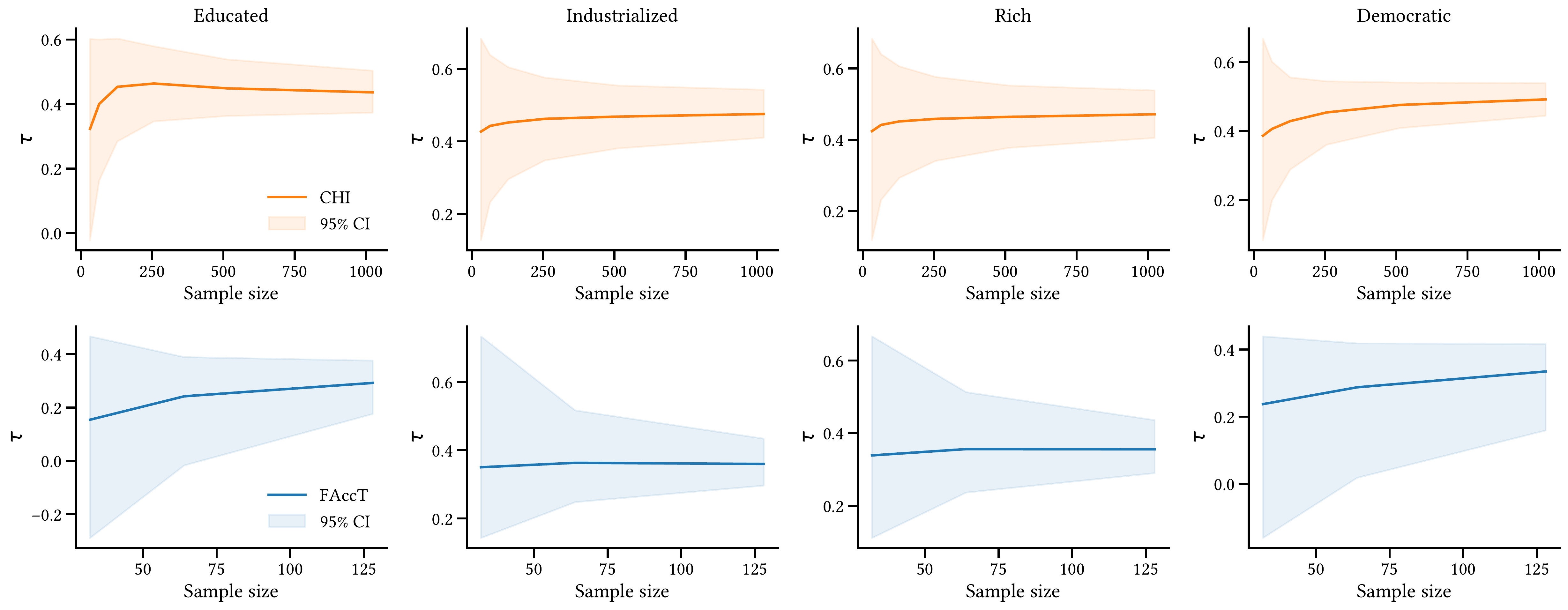}
    \caption{To see the effect of sample size, we did bootstrap sampling with different number of papers using a power of 2 (i.e., 16, 32, \ldots, maximum number of papers for the respective conference). The result suggests that FAccT is less ``EIRD'' than CHI.}
    \label{fig:sample_size}
\end{figure*}

\noindent \textbf{Sample Size.}
Tables~\ref{table:western} and \ref{table:weird_index} show that FAccT has a larger proportion of papers with exclusively Western participants compared to CHI but is less ``EIRD'' (i.e., the Educated, Industrialized, Rich, and Democratic scores are lower). However, we expected a positive correlation between Western and EIRD variables. A possible explanation of this counter-intuitive result is that EIRD scores are measured using Kendall tau rank correlation, which is heavily based on the set of sampled countries. However, we tested whether the difference between the two conferences would impact our results and that was not the case (Table~\ref{table:weird_intersection}).
The two conferences have different sets of sampled countries: FAccT from 50 unique countries, whereas CHI from 93 unique countries. Therefore, to test whether EIRD scores are associated with the difference in sampled countries, we computed the Kendall tau rank correlation for each conference using the intersection of the two sets of countries (Table~\ref{table:weird_intersection}). We found that FAccT is still less ``EIRD'' compared to CHI. We further explored this difference by bootstrapping the number of papers from each conference using a power of 2 (i.e., 16, 32, \ldots, maximum number of papers for the respective conference). We sampled by replacement from the pool of papers 10,000 times for each sample size and computed the EIRD scores each time. This approach allowed us to quantify the uncertainty associated with the scores. Again, FAccT is less ``EIRD'' than CHI (Figure~\ref{fig:sample_size}). 

\begin{table}
    \centering
    \caption{We compared the effect of using off-the-shelf datasets (e.g., those from UCI Machine Learning Repository) and author-collected datasets (e.g., interviews and surveys) to the Educated, Industrialized, Rich, and Democratic scores by randomly shuffling the two types of datasets and calculating the difference in EIRD scores ($\Delta \tau$). We found that the actual differences fall within the 95\% confidence interval from the shuffling, indicating that there is insufficient evidence to suggest that these differences are significant.}
    \label{tab:original}
    \begin{tabular}{lcc}
        \toprule
        Variable & $\Delta \tau$ & 95\% CI \\
        \midrule
        Educated & -0.03 & [-0.52, 0.15] \\
        Industrialized & 0.09 & [-0.20, 0.26] \\
        Rich & 0.06 & [-0.21, 0.28] \\
        Democratic & 0.02 & [-0.51, 0.18] \\
        \bottomrule
    \end{tabular}
\end{table}

\noindent \textbf{Off-the-Shelf Datasets vs. Author-Collected Datasets.} 
Off-the-shelf datasets are publicly available (e.g., UCI Machine Learning Repository). In contrast, author-collected datasets are those collected through interviews, surveys, or workshops. We examined whether overused datasets may explain the bias in the sample population by comparing studies that used off-the-shelf and author-collected datasets. Over-reliance on certain datasets may lead to bias by repeatedly using participants from the same country. However, there are also widely-used datasets that include samples from multiple countries such as the drug consumption~\cite{uci2017ml} and the VoxCeleb 1~\cite{nagrani17interspeech}. Additionally, while the majority of participants in FAccT papers are from the U.S. due to the use of datasets such as the Adult, the COMPAS recidivism, and the ACS, these studies typically focus on specific subgroups within these datasets such as native Americans, Hispanic, and Black communities~\cite{celis2019fairness, sikdar2022getfair, grabowicz2022marry}. Other off-the-shelf datasets used in FAccT papers also include participants from multiple countries such as FairFace~\cite{karkkainen2021fairface} in~\cite{wolfe2022markedness} and Labeled Faces in the Wild~\cite{huang2007lfiw} in~\cite{buet2022towards, ghadiri2021socially}. However, since the countries of origin for the participants in these datasets are not specified, these papers were not included in the analysis.

\begin{figure}
    \centering
    \includegraphics[width=\columnwidth]{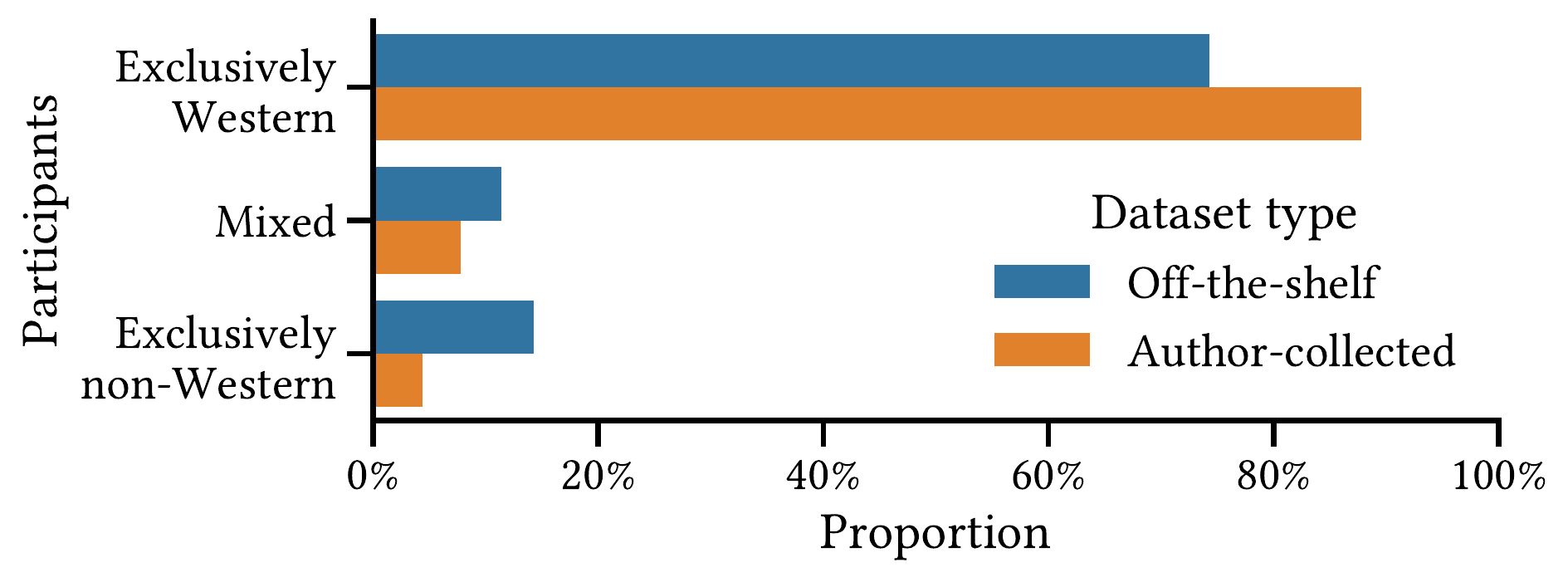}
    \caption{How types of datasets are associated with the proportion of sampled Western participants in FAccT.}
    \label{fig:western-origin}
\end{figure}

\begin{figure*}[tbp]
    \centering
    \includegraphics[width=\textwidth]{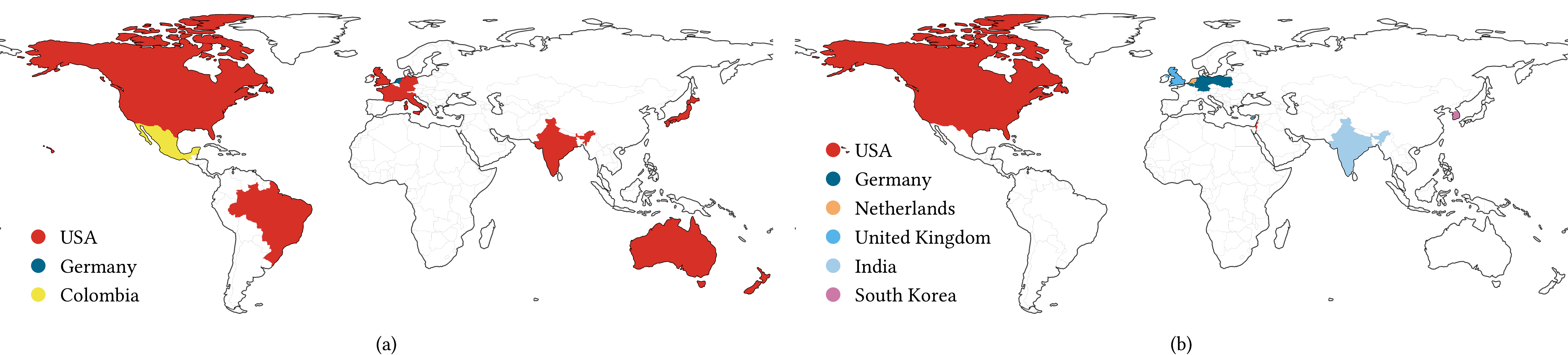}
    \caption{The maps are colored based on the majority of study participants' country of origin. FAccT authors mainly used off-the-shelf datasets from the U.S (a), while datasets collected by the authors (e.g., interviews or surveys) have a more localized representation (b).}
    \label{fig:world-map-datasets}
\end{figure*}

From the 128 analyzed papers, we found that 90 (70.3\%) of them were only using off-the-shelf datasets, 35 (27.3\%) were using author-collected datasets, and 3 (2.4\%) were using a mix of both. The top three off-the-shelf datasets were the Adult Income~\cite{uci2017ml} (23 papers, 18\%), the COMPAS Recidivism~\cite{larson2023compas} (21 papers, 16.4\%), and the German credit~\cite{hofmann1994german} (13 papers, 10.2\%). Our initial assumption was that there would be no difference between the two types of datasets in terms of WEIRD-ness (excluding the 3 papers with mixed datasets). To test the EIRD part, we randomly shuffled the labels from the two sets, and found no statistically significant difference (Table~\ref{tab:original}). 
A chi-square test of independence was performed to examine the relation between the Western score and the type of datasets (i.e., off-the-shelf vs. author-collected), and found no statistically significant association between the two variables (${\chi}^2 (2, N = 125) = 4.32, p = 0.16$). However, Figure~\ref{fig:western-origin} shows that the author-collected datasets are slightly less Western. Separating papers based on the majority of participants' country of origin also shows subtle differences. FAccT authors mainly used off-the-shelf datasets from the U.S., while datasets collected by authors (e.g., interviews or surveys) have a more localized representation (Figure~\ref{fig:world-map-datasets}).

\begin{table}[tbp]
    \centering
    \caption{The number of unique author affiliation countries is not associated with any of the ``EIRD'' variables for FAccT, while there is an inverse correlation for CHI. $\rho$ is the Pearson correlation coefficient.} 
    \label{tab:affiliations}
    \begin{tabular}{lcc|cc}
        \toprule
        {} & \multicolumn{2}{c}{\textbf{FAccT}} & \multicolumn{2}{c}{\textbf{CHI} (\citeauthor{linxen2021weird})} \\
        Variable & $\rho$ & 95\% CI & $\rho$ & 95\% CI \\
        \midrule
        Educated & -0.06 & [-0.24, 0.10] & -0.22 & [-0.30, -0.14] \\
        Industrialized & -0.07 & [-0.25, 0.10] & -0.21 & [-0.29, -0.14] \\
        Rich & -0.06 & [-0.25, 0.10] & -0.23 & [-0.30, -0.16] \\
        Democratic & 0.03 & [-0.13, 0.20] & -0.10 & [-0.17, -0.03] \\
        \bottomrule
    \end{tabular}
\end{table}

\begin{table}[tbp]
    \centering
    \caption{The Educated, Industrialized, Rich, and Democratic (EIRD) scores differences ($\Delta \tau$) between papers with authors from a single country and multiple countries. Positive values of $\Delta \tau$ indicate that papers with authors from a single country are sampling from more EIRD countries. By randomly shuffling the two types of papers and calculating the differences in scores, we found that the actual differences fall within the 95\% confidence interval from the shuffling, indicating that these differences are not strong.} 
    \label{tab:cross-country-diff}
    \begin{tabular}{lcc|cc}
        \toprule
        {} & \multicolumn{2}{c}{\textbf{FAccT}} & \multicolumn{2}{c}{\textbf{CHI} (\citeauthor{linxen2021weird})} \\
        Variable & $\Delta \tau$ & 95\% CI & $\Delta \tau$ & 95\% CI \\
        \midrule
        Educated & 0.43 & [-0.22, 0.59] & 0.11 & [-0.17, 0.12] \\
        Industrialized & 0.09 & [-0.37, 0.23] & 0.09 & [-0.12, 0.18] \\
        Rich & 0.13 & [-0.33, 0.25] & 0.09 & [-0.12, 0.18] \\
        Democratic & 0.28 & [-0.17, 0.55] & 0.02 & [-0.07, 0.16] \\
        \bottomrule
    \end{tabular}
\end{table}

\noindent \textbf{Author Affiliation Countries.} We examined whether the geographical differences in authors' affiliations explain the sample representativeness of research studies. We wanted to understand whether authors from multiple countries are associated with the research paper being less WEIRD (inspired by discussions and proposals made by \citet{linxen2021weird}). The geographical location of an author's affiliation can influence the population being studied. For example, suppose a study is conducted by researchers affiliated with institutions in the U.S. In that case, it is more likely that the sample will be composed of participants from the U.S. Conversely, if a study is conducted by researchers affiliated with institutions in Asia, it is more likely that the sample will be composed of participants from Asia. This is important to consider as it can lead to a lack of diversity in the sample and potential biases in the results. Furthermore, the difference in countries can also influence the type of data collection methods used and the type of questions being asked, which can also impact the sample's representativeness. Thus, we used Pearson correlation coefficients to measure the relationship between the number of countries that authors are affiliated with and the EIRD variables.

The first author manually extracted authors' countries of affiliation from the 128 papers, using the affiliations reported in the papers. For authors with multiple affiliations, their first affiliation was used. Out of 519 authors, we were unable to determine the countries of 11 (2.2\%), thus excluding them from this analysis. All analyzed papers ($n = 128$) have at least one author with their country listed, and the average number of authors is equal to $4.1$ ($ \sigma = 2.8 $). We found that authors represent 20 unique countries (65\% Western), and 75\% of papers come from authors from a single country. Furthermore, we computed the Pearson correlation coefficients, one for each of the EIRD variables (Table~\ref{tab:affiliations}). We found no significant association between the number of countries where the authors are affiliated with, and the EIRD variables.

To further explore the variations in EIRD scores, we divided the 128 papers into two mutually exclusive groups based on whether the authors were from one ($n=96$) or multiple ($n=32$) countries. We used a shuffling method to ensure the significance of the differences by re-arranging the paper labels 10,000 times, and calculating the differences in EIRD scores each time. Despite the noticeable differences in the Educated and Democratic scores, the results did not show statistical significance (Table~\ref{tab:cross-country-diff}). Given the number of analyzed papers that have been further divided into the two groups, these results may be due to a lack of statistical power (see \S\ref{sec:limitation}). To partly address that, we conducted a similar analysis by using CHI data \cite{linxen2021weird}. Despite coming from 70 unique countries, the authors of the 1076 papers tended to collaborate with authors in the same country; only 23\% of papers were written by authors from at least two countries.

\begin{figure}[htbp]
    \centering
    \includegraphics[width=\columnwidth]{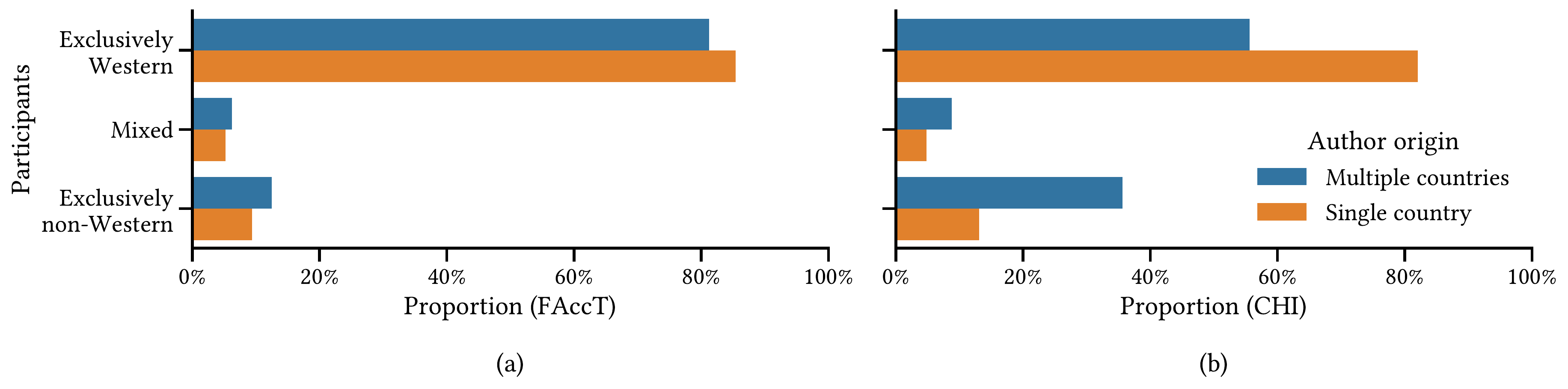}
    \caption{How author countries are associated with the proportion of sampled Western participants for (a) FAccT and (b) CHI. Papers produced by authors from multiple countries are more likely to sample from non-Western countries. The difference is more apparent on CHI, possibly due to the sample size difference between FAccT and CHI.}
    \label{fig:western}
\end{figure}

The Pearson correlation coefficients indicate that there is a negative correlation between the number of unique author affiliation countries and the EIRD variables (Table~\ref{tab:affiliations}). However, the differences in EIRD scores ($\Delta \tau$) observed using the shuffling method are only statistically significant ($p < 0.05$) for the Educated score (Table~\ref{tab:cross-country-diff}). This means that when CHI authors come from diverse countries, it is less likely for the resulting paper to be focused on Educated samples. We also examined the differences between FAccT and CHI for the Western variable based on the origin of the authors (i.e., whether they come from a single or multiple countries). Papers written by authors from multiple countries are less likely to have participants from exclusively Western countries (Figure~\ref{fig:western}). However, the difference in proportion is more apparent in CHI than in FAccT. This finding corroborates what was suggested by \citet{linxen2021weird} to ``foster collaborations across Western and non-Western countries'' to make CHI less WEIRD.

\section{Discussion}
\label{sec:discussion}
By analyzing 128 papers published between 2018 and 2022 at the FAccT conference, we studied the extent to which the conference relies on WEIRD samples and overused datasets. We found that 84\% of FAccT papers are based exclusively on Western participants (representing less than 12\% of the world's population), 7\% exclusively on non-Western participants, and 9\% a mix of both. Compared to CHI, the leading conference on human factors in computing, FAccT papers draw less from Educated, Industrialized, Rich, and Democratic samples but more from Western ones. Comparing the WEIRD-ness between CHI and FAccT is crucial for identifying potential gaps in research practices, methodologies, and datasets, given that FAccT is a relatively new conference with limited representation. While the lack of statistically significant differences does not refute the presence of WEIRD-ness, the comparison serves as an initial step towards recognizing disparities and promoting a more inclusive research community. Moreover, insights from the CHI conference offer foresight into future issues, providing valuable guidance for researchers toward more inclusive practices. Thus, the following subsections are based on results from both FAccT and CHI.

\subsection{Implications}
\label{sec:implications}
From a \textbf{theoretical standpoint}, our work contributes to the growing body of literature concerning WEIRD research, situated more broadly in the literature of Critical Computing~\cite{comber2020announcing}. We echo the criticisms raised by \citet{laufer2022four} that the FAccT community is Western- and U.S.-centric, including the ``\textit{set of values and modes of discourse}.'' A similar finding was also reported on \citet{van2023methodology}'s study comparing the same two conferences, albeit the study was conducted on a smaller sample size. Our study embarks on this strand of research by not only surfacing the WEIRD-ness of both conferences but also by uncovering their differences. In the broader Responsible AI context, questions remain to be quantitatively answered. Of prime importance is the so-called phenomenon of ``\textit{fair-washing}.'' As \citet{laufer2022four}'s described it, this phenomenon concerns the ``\textit{narrow notions of fairness [that are used] to condone existing practices},'' linking it with unregulated corporate influence on the scholarly discourse; industry interests may not necessarily be aligned with those of scholars, adding, or even amplifying biases, and minimally helping those affected by algorithmic systems. Waldman also points out that the discourse on privacy is heavily influenced by industry and its benefits~\cite{waldman2021industry}. 

From a \textbf{practical standpoint}, our findings speak to the need to take steps to make conferences less WEIRD. Here, we echo statements made by \citet{sambasivan2021everyone} for ``\emph{incentivizing data excellence as a first-class citizen of AI, resulting in safer and more robust systems for all.}'' What could be learned from best practices across conferences, and how organizing committees could embrace these practices? Next, we provide suggestions to alleviate FAccT's WEIRD-ness:

\noindent \textbf{Mandatory reproducibility statement}: FAccT could prioritize reproducibility by promoting the replication and extension of findings and encouraging the sharing of data and materials. As a case in point, the Web (WWW) conference has enforced for reproducibility statement~\cite{www2022}. Such a statement not only allows for generalizable and reproducible findings but is also a way to help surface and mitigate the WEIRD-ness of a conference. Merely promoting research reproducibility may not adequately address WEIRD-ness if biased participant selection is the underlying issue. Nevertheless, when coupled with studies that highlight the importance of diverse participation in research and the provision of datasets containing non-WEIRD participants, it may result in the development of more varied off-the-shelf datasets for benchmarking purposes. With this approach, we foresee community-wide adoption of these datasets~\cite{wilkinson2016fairdata}.


\noindent \textbf{Mandatory data statement}: FAccT could implement a mandatory data statement policy, requiring authors to report the geographic breadth of their participant samples. This also allows conference attendees to evaluate the generalizability of the research and understand its limitations. A \textbf{visualization tool} could do a real-time visualization of these points to provide a quick overview of datasets' characteristic to organizers, authors, and attendees.

\noindent \textbf{Champion for author diversity across FAccT and other venues}: FAccT could prioritize diverse authorship by actively seeking and encouraging submissions from researchers from different countries and cultures. Limiting our analysis on FAccT may have resulted in missing this important point due to the small sample size. However, results from our CHI analysis (\S\ref{sec:rq2}) show promises that such diversity can help make conferences less WEIRD (Figure~\ref{fig:western}). One idea could be to encourage collaboration using \textbf{speed dating} for attendees to encourage collaboration between authors that may not meet outside the conference. Attendees could randomly get assigned to tables for 5 minutes to introduce each other and give a short description of their to spark creativity and possible future collaborations.

\noindent \textbf{Encourage cross-country online research methods to recruit from a diverse pool of participants}: FAccT could create sessions or workshops focused on online research methods and their potential for studying geographically diverse samples. This could include training and support for researchers on effectively using these methods and overcoming any challenges they may face. FAccT could collaborate with other academic venues to incentivize data excellence by making data statements and datasheets for datasets mandatory for authors submitting their work.

\subsection{Limitations and Future Work}
\label{sec:limitation}
Our work has three limitations that call for future research efforts. First, we limit our analyses to five years due to the FAccT's starting date, and the two sets of dates of the two conferences are not aligned because the CHI conference has a longer tenure. As the community grows, future studies could replicate our methodology, allowing for a broader comparison between the two communities (i.e., CHI and FAccT), accounting for more years of proceedings to be included. Along the same lines, the FAccT proceedings tend to attract a relatively larger number of studies concerned with datasets compared to CHI. Again, any differences attributed to that would be possible to surface. The findings in \S\ref{sec:rq2} indicate insufficient evidence concluding that there is a significant difference in EIRD scores between datasets collected off-the-shelf and by the authors or between papers produced by authors from one country vs. multiple countries. However, these results should be interpreted with the caveat that the analysis was based on 128 papers, which had to be divided into these two groups. The limited sample size is, therefore, likely to decrease statistical power, leading to results with a low probability of accurately detecting a real effect or results influenced by both random and systematic errors. Nevertheless, by replicating \citet{linxen2021weird}'s dataset, we still observed statistically significant differences to some extent using a dataset of similar nature but nearly ten times larger in size. Second, we analyzed 30.8\% of the papers due to the lack of relevant dataset information for the rest. We attribute this to not having a unified way of sharing datasets in FAccT. This speaks to the need for incentivizing data excellence and sharing, allowing for reproducibility (\S\ref{sec:implications}). Future research should also replicate this methodology in other conferences, establishing a single WEIRD metric (like impact factor for a conference's or journal's importance) for research diversity. By embracing such a metric, we foresee research diversification and focus on non-WEIRD populations. Third, this work considered only the WEIRD variables. Future research may well study datasets commonly used in the AI community through the lens of gender, class, and race, as well as a combination of demographics that can get lost when these factors are treated separately and as concrete groups~\cite{schlesinger2017intersectional}; a person could be part of multiple marginalized groups, which can result in compounding effect (e.g., black woman or an immigrant with a disability). Moreover, categorizing the world into uniform ``civilizations'' as defined by \citet{huntington2000clash} is an oversimplification, ignoring the complex and nuanced nature of cultures that are evolving worldwide (including Western nations). Each culture is diverse and continually changing, shaped by unique combinations of historical, geographical, social, and economic factors. Recognizing and valuing the complexity and diversity of cultures is essential for inclusive and representative research. When examining the samples utilized in research studies, it may also appear that a significant proportion of them are from regions that are not WEIRD. However, upon closer inspection of factors such as the authors' background, institutions, methodologies, and results interpretation, it may still be apparent that a given research is heavily influenced by Western culture. Future studies could also focus on
various perspectives, such as discussing counter-algorithm movements~\cite{benjamin2022fuck} or understanding the societal impact of algorithms~\cite{krafft2021advocate}.
\section{Conclusion}
\label{sec:conclusion}
Of 128 papers published at FAccT between 2018 and 2022, 84\% were based on exclusively Western participants. Compared to the CHI conference, we found that FAccT is less EIRD. We did not find conclusive answers by exploring several hypotheses about the reasons for such differences. These differences are neither attributed to the types of datasets used (off-the-shelf vs. author-collected) nor to cross-affiliation collaborations. We encourage the FAccT community to increase sample diversity and address potential biases in the findings presented at the conference.

\newpage



\bibliographystyle{ACM-Reference-Format}
\bibliography{main}


\end{document}